\documentclass[sigconf]{acmart}

\AtBeginDocument{%
  \providecommand\BibTeX{{%
    \normalfont B\kern-0.5em{\scshape i\kern-0.25em b}\kern-0.8em\TeX}}}

\copyrightyear{2024}
\acmYear{2024}
\setcopyright{acmlicensed}\acmConference[KDD '24]{Proceedings of the ACM SIGKDD Conference on Knowledge Discovery and Data Mining 2024}{August 25--29, 2024}{Barcelona, Spain}
\acmBooktitle{Proceedings of the ACM SIGKDD Conference on Knowledge Discovery and Data Mining 2024 (KDD '24), August 25--29, 2024, Barcelona, Spain}
\acmDOI{xx.xxxx/xxxxxxx.xxxxxxx}
\acmISBN{xxx-x-xxxx-xxxx-x/xx/xx}

\usepackage{amsfonts}
\usepackage{algorithm}
\usepackage{algcompatible}

\usepackage{graphicx}
\usepackage{float}
\usepackage{multirow}
\usepackage{balance}

\begin{document}

\title{Diffusion-Based Cloud-Edge-Device Collaborative Learning for Next POI Recommendations}

\author{Jing Long}
\email{jing.long@uq.edu.au}
\orcid{1234-5678-9012}
\affiliation{%
 \institution{The University of Queensland}
 \city{Brisbane}
 \state{QLD}
 \postcode{4072}
 \country{Australia}}

\author{Guanhua Ye}
\email{rex.ye@dncc.tech}
\affiliation{%
 \institution{Deep Neural Computing Company Limited}
 \city{Shenzhen}
 \state{Guangdong}
 \postcode{518000}
 \country{China}}

 \author{Tong Chen}
\email{tong.chen@uq.edu.au}
\affiliation{%
 \institution{The University of Queensland}
 \city{Brisbane}
 \state{QLD}
 \postcode{4072}
 \country{Australia}}

\author{Yang Wang}
\email{yangwang@hfut.edu.cn}
\affiliation{%
 \institution{Hefei University of Technology}
 \city{Hefei}
 \state{Anhui}
 \postcode{230000}
 \country{China}}

\author{Meng Wang}
\email{eric.mengwang@gmail.com}
\affiliation{%
 \institution{Hefei University of Technology}
 \city{Hefei}
 \state{Anhui}
 \postcode{230000}
 \country{China}}
 
\author{Hongzhi Yin*}
\email{h.yin1@uq.edu.au}
\affiliation{%
 \thanks{*Corresponding author}
 \institution{The University of Queensland}
 \city{Brisbane}
 \state{QLD}
 \postcode{4072}
 \country{Australia}}

\renewcommand{\shortauthors}{Jing Long et al.}

\begin{abstract}
The rapid expansion of Location-Based Social Networks (LBSNs) has highlighted the importance of effective next Point-of-Interest (POI) recommendations, which leverage historical check-in data to predict users' next POIs to visit. Traditional centralized deep neural networks (DNNs) offer impressive POI recommendation performance but face challenges due to privacy concerns and limited timeliness. In response, on-device POI recommendations have been introduced, utilizing federated learning (FL) and decentralized approaches to ensure privacy and recommendation timeliness. However, these methods often suffer from computational strain on devices and struggle to adapt to new users and regions. This paper introduces a novel collaborative learning framework, Diffusion-Based Cloud-Edge-Device Collaborative Learning for Next POI Recommendations (DCPR), leveraging the diffusion model known for its success across various domains. DCPR operates with a cloud-edge-device architecture to offer region-specific and highly personalized POI recommendations while reducing on-device computational burdens. DCPR minimizes on-device computational demands through a unique blend of global and local learning processes. Our evaluation with two real-world datasets demonstrates DCPR's superior performance in recommendation accuracy, efficiency, and adaptability to new users and regions, marking a significant step forward in on-device POI recommendation technology.
\end{abstract}

\ccsdesc[500]{Information systems~Recommender systems}

\keywords{Point-of-Interest Recommendation; On-Device POI Recommendations; Diffusion Models}

\maketitle
\section{Introduction}
The emergence of Location-Based Social Networks (LBSNs), such as Foursquare and Weeplace, has improved the way we interact with our surroundings. These platforms, accumulating vast amounts of historical check-in data, have become fertile ground for developing Point-of-Interest (POI) recommendation systems. Given the powerful computational capabilities of servers, centralized deep neural networks (DNNs) based on graph embedding \cite{li2021discovering,2021Graph,yin2016spatio} and attention mechanisms \cite{2020Geography,2021STAN} demonstrate impressive performance in POI recommendations. Unfortunately, due to increasing concerns regarding privacy and the location-sensitive nature of POI recommendations, users are becoming increasingly cautious and even reluctant to upload their check-in data, thereby impacting the recommendation quality \cite{long2024physical,zhang2022pipattack}. Apart from this, recommendations in centralized services are computed upon request and then transmitted to user devices, making the service timeliness highly dependent on network quality \cite{long2023decentralized}. Thus, on-device POI recommendations have emerged, aimed at mitigating the limitations of centralized paradigms \cite{yin2024device}. That is, each user locally hosts a lightweight recommendation model that generates personalized recommendations without sharing sensitive data, which also warrants responsiveness. 

Being a widely recognized approach under this paradigm, federated learning (FL) based POI recommendations (e.g., \cite{2021PREFER}) centrally collect and aggregate locally trained models, as well as redistributing the aggregated model to all users. However, all users sharing the same model may hurt the minority groups and impair the recommendation quality of these users. To achieve a higher degree of personalization, instead of aggregating all user models, some federated POI recommenders \cite{2020xw,2021rao,imran2023refrs,wang2017location} group similar users and perform aggregation within groups. Further remedy is proposed in decentralized POI recommenders \cite{long2023decentralized,long2023model}, where users can directly engage in collaborative learning with their neighbors in a device-to-device manner, allowing more personalization of learned on-device models. Unfortunately, the aforementioned federated and decentralized frameworks suffer from two major limitations. On the one hand, they require full device engagement during training or updating, whether in collaboration with the cloud or other devices, which heavily burdens on-device computational resources. On the other hand, they face challenges in transferability, as they must learn patterns for new users and regions from the ground up.

To this end, we propose a fast-adapting on-device POI recommendation framework, namely Diffusion-Based Cloud-Edge-Device Collaborative Learning for Next POI Recommendations (DCPR). DCPR adopts the diffusion model as its primary building block, which has drawn significant attention due to its substantial success in various fields like computer vision (CV), natural language processing (NLP), sequential recommendations, and others \cite{avrahami2022blended,li2022diffusion,ramesh2022hierarchical,yang2023diffusion}. Leveraging its advantages in distributed generation and diverse representations, we believe the diffusion model is highly suitable to bridge the above gap. More intuitively, the proposed framework consists of three layers, including cloud server, edge server, and device. Initially, a centrally hosted global diffusion model is trained to learn category-level movement patterns. Since the training data (i.e., category sequences) does not involve sensitive geographical locations, it is easily collected in LBSNs.

\begin{figure}
	\includegraphics[width=\linewidth]{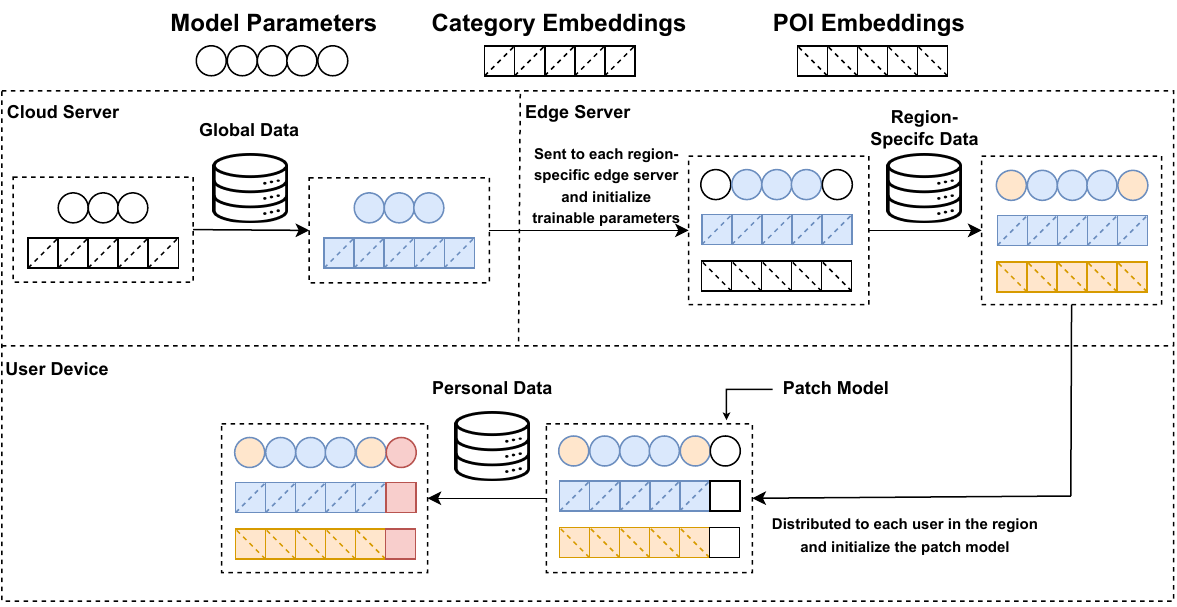}
        \vspace{-1.0em}
	\caption{The overview of our proposed DCPR.}
	\label{overview} 
\end{figure}

Subsequently, the well-trained global model is sent to all region-specific edge servers, and endowed with the ability to capture region-specific preferences. This is achieved by each region-specific edge server modifying the global model with POI sequences in this region. The training data for each region comes from published de-identified check-in sequences. Finally, each edge server distributes the region-specific model to all users within this region, which is further fine-tuned locally by personal data. To avoid impairing the inherent generation capabilities of the region-specific model, an additional patch model is inserted and updated with personal data to reflect the user's personal preferences. An acceleration is further adopted to speed up the inference process, which is a significant drawback of the standard diffusion model. With the cloud-edge-device architecture, DCPR significantly reduces the burden of the on-device computational resource, having capabilities to provide POI recommendations effectively and efficiently. Meanwhile, such progressively personalized architecture is highly transferable as it can rapidly adapt to new regions and new users. In a nutshell, we summarize our contributions as follows:

\begin{itemize}
\item To the best of our knowledge, we are the first to bridge the gap between the diffusion model and on-device POI recommendations and propose a fast-adapting on-device POI recommendation framework, namely DCPR, aimed at providing personalized POI recommendations efficiently.
\item DCPR consists of three layers including cloud server, edge server, and device, where the latter is progressively built on the former, and thus, it can fit new regions and users quickly. To speed up the on-device inference process, we further design an acceleration strategy,  significantly reducing inference time.
\item We evaluate DCPR with two real-world datasets, and demonstrate its effectiveness. The experimental results highlight superior accuracy, efficiency, and transferability.
\end{itemize}
\vspace{-1.0em}
\section{Preliminaries}\label{sec:prelim}
In this section, we list key notations used throughout this paper, outline our primary task, and introduce the standard diffusion model.

\vspace{-1.0em}
\subsection{Notations}
We denote the sets of users $u$, POIs $p$ and categories $c$ as
$\mathcal{U}$,
$\mathcal{P}$, 
$\mathcal{C}$, 
respectively. Each POI $p\in \mathcal{P}$ is associated with a category tag (e.g., entertainment or restaurant) $c_p\in \mathcal{C}$ and coordinates $(lon_p,lat_p)$.

\textbf{Definition 1: Check-in Sequence}. A check-in activity of a user indicates a user $u\in \mathcal{U}$ has visited POI $p\in \mathcal{P}$ at timestamp $t$. By sorting a user's check-ins chronologically, a check-in sequence contains $M_u$ consecutive POIs visited by a user $u$, denoted by $\mathcal{X}(u)=\{p_1, p_2,...,p_{M_u}\}$. Each personal check-in sequence $\mathcal{X}(u)$ is stored on the corresponding personal device. 

\textbf{Definition 2: Category Sequence}. A category sequence substitutes all POIs in the check-in sequence $\mathcal{X}(u)$ with their associated category tags, indicated as $\mathcal{X}^c(u)=\{c_1,c_2,...,c_{M_u}\}$.

\textbf{Definition 3: Global POI Category Sequence Dataset}. The global semantic dataset $\mathcal{D}_g = \{{\mathcal{X}_z^c}\}_{z=1}^Z$ consists of $Z$ anonymized categorical sequences.  The global POI category sequence dataset is stored on the cloud server. 

\textbf{Definition 4: Region}. A region $r$ refers to a geographic segment providing additional context about the POIs it encompasses.  We do not assume specific region division methods,  although we adopt $k$-means clustering \cite{1967Some} to derive a set of regions $\mathcal{R}$ following \cite{long2023model}.  Other predefined functional regions, such as city districts or suburbs, can also work in our proposed framework. 

\textbf{Definition 5: Region-Specific Dataset}. The region-specific dataset $\mathcal{D}_r=\{\mathcal{X}_v\}_{v=1}^{V_r}$ for a region $r$ includes $V_r$ anonymized check-in sequences. Each region possesses its unique dataset, encompassing check-in activities exclusively within region $r$.  Each region-specific dataset $\mathcal{D}_r$ is stored on a region-based edge server. 

\subsection{Task: On-Device Next POI Recommendation}
With the cloud-edge-device architecture, DCPR owns a cloud server and multiple edge servers where each region is assigned an edge server. Then, the functions of cloud server, edge server, and user device are as follows:
\begin{itemize}
\item \textbf{Cloud Server.} The cloud server initially develops a global diffusion network $\Theta_g$ with the global dataset $\mathcal{D}_g$. Subsequently, $\Theta_g$ is sent to all region-specific edge servers. 
\item \textbf{Edge Server.} After receiving the global network $\Theta_g$, each region-specific edge server modifies it with the region's check-in sequences $\mathcal{D}_r$. The edge server then distributes the customized region-specific model $\Theta_r$ to all users within the region.
\item \textbf{User Device.} Each user $u$ receive the region-specific model $\Theta_r$ and further refine it to create a personalized model $\Theta_u$ with personal data $\mathcal{X}_u$.
\end{itemize}
Under this construction, we aim to develop a performant local model for each user, capable of providing a ranked list of potential POIs for the next visit.

\subsection{Standard Diffusion Model}\label{sdm}
Before introducing our model, we briefly describe the standard diffusion model as the preliminary knowledge. The \textbf{Diffusion Phase} progressively transforms the initial representation $x_0$ into pure Gaussian noise via a Markov Chain $(x_0\rightarrow x_1\rightarrow x_2 \rightarrow ... \rightarrow x_{t-1}\rightarrow x_t \rightarrow ... \rightarrow x_{T-1} \rightarrow x_T)$, where $T$ denotes the maximum diffusion step. More specifically, the relationship between $x_t$ and $x_{t-1}$ is formulated as:
\begin{equation}
    x_t = \sqrt{1-\beta_t}x_{t-1} + \beta_t\epsilon,
\end{equation}
where $\epsilon\sim \mathcal{N}(0,I)$ which is a standard normal distribution, and $\beta_t$ controls noise level at the diffusion step $t$. Recall that the diffusion process aims to make $x_0$ converge towards a standard normal distribution, $\beta_t$ increases with the growth of $t$. Normally, the value of $\beta_t$ is generated from a pre-defined noise schedule, while common noise schedules include square-root schedule \cite{li2022diffusion}, cosine schedule \cite{ho2020denoising}, and linear schedule \cite{nichol2021improved}. In this paper, we adopt the square-root schedule, and $\beta_t$ is defined as:
\begin{equation}
    \beta_t = \sqrt{t/T + w},
\end{equation}
where $w$ is a small constant corresponding to the starting noise level. Inspired by \cite{ho2020denoising}, $x_t$ can also be derived directly from the original target category embedding $x_0$, where the relationship is defined as:
\begin{equation}
    x_t = \sqrt{\overline{\alpha}_t}x_0 + \sqrt{1-\overline{\alpha}_t}\epsilon,
\end{equation}
where $\epsilon\sim \mathcal{N}(0,I)$ and
\begin{equation}
    \sqrt{\overline{\alpha}_t}=\prod^t_{s=1}\alpha_s,
\end{equation}
where $\alpha_s = 1-\beta_s$. In this way, a vast amount of training samples are obtained to train a network $\Theta$, having the capability to estimate the original presentation $x_0$ given its noised version $x_t$.

The \textbf{Reverse Phase} denoises the pure Gaussian noise $x_T$ to approximate the initial representation $x_0$ in an iterative manner $(x_T\rightarrow x_{T-1} \rightarrow ... \rightarrow x_t\rightarrow x_{t-1} \rightarrow ... \rightarrow x_1 \rightarrow x_0)$, which is precisely opposite to the diffusion process. Formally, $x_{t-1}$ is obtained from $x_t$ by:
\begin{equation}
    x_{t-1} = \frac{\sqrt{\alpha_t}(1-\overline{\alpha}_{t-1})x_t+\sqrt{\overline{\alpha}_{t-1}}(1-\alpha_t)\hat{x}_0+(1-\alpha_t)(1-\overline{\alpha}_{t-1})\epsilon}{1-\overline{\alpha}_t},
\end{equation}
where $\hat{x}_0 = \Theta(x_t)$, and $\epsilon\sim \mathcal{N}(0,I)$.

\section{Methodology}
In this section, we formally introduce the design of DCPR, with an overview provided in Figure \ref{overview}. The framework consists of three stages: 
\textbf{(1) Development of a Global Diffusion Model:} This stage involves creating a model adept at encapsulating category-level inclinations on the cloud server.
\textbf{(2) Training of Region-Specific Models:} The global model is tailored to each region by incorporating check-in sequences pertinent to that region on the region-specific edge server, thereby creating models attuned to regional dynamics.
\textbf{(3) Local Finetuning for Personalization:} In the final stage, the region-specific models undergo local refinement to generate personalized models. This ensures highly accurate and tailored on-device POI recommendations for individual users.

Selecting the diffusion model as the backbone in our proposed framework for on-device POI recommendations is driven by two key considerations. Firstly, the diffusion model's architecture excels at managing complex data \cite{ho2020denoising}, an essential feature for POI recommendations that demand a thorough grasp of the nuanced and dynamic aspects of POI sequences. This alignment greatly enhances the model's generalization capabilities across different scenarios, leading to precise recommendations. Secondly, the diffusion model's strengths in distributed generation and its capability to effectively capture a wide range of representations make it uniquely suited to swiftly and accurately adapt to new regions and user profiles, benefiting its transferability. 

\subsection{Global Diffusion Model}
The proposed framework begins with the training of a global diffusion model on the cloud server, aimed at learning the patterns of category-level movements.

\subsubsection{Diffusion Phase}
Formally, the global diffusion model is designed to construct the next category embedding $\textbf{e}^c_{M+1}$ from Pure Gaussian noise $x_T$, where $x_T \sim\mathcal{N}(0, I)$, conditional on the historical category sequence $\mathcal{X}_c=\{c_1,c_2,...,c_{M}\}$. Here, the next category embedding $\textbf{e}^c_{M+1}$ is also known as the initial target representation $x_0$. Following the standard diffusion algorithm, we progressively add noise into the target category embedding $x_0$ through the diffusion phase, and represent the noise-altered target category at step $t$ as $x_t$. This process is leveraged to generate samples for training a network $\Theta_g$, which takes as input the noised target category representation $x_t$ and the category sequence $\mathcal{X}_c$. The output is the estimated target category representation $\hat{x}_0$, aiming to approximate the true target category embedding $x_0$. The comprehensive design of the network $\Theta_g$ will be introduced in the following section. Since the task of the proposed DCPR is to offer on-device POI recommendations, the global network $\Theta_g$ serves merely as a semi-finished model and does not engage in the reverse phase (inference) solely. We will describe the personal inference process later.

\subsubsection{Attention-based Network}\label{global attention network}
As previously mentioned, the core objective of the network is to reconstruct the target category embedding $x_0$ given its noised representation $x_t$ and the historical category sequence $\mathcal{X}_c$, denoted as: 
\begin{equation}
    \hat{x}_{0}=\Theta_g(x_t,\mathcal{X}_c),
\end{equation}
where $\hat{x}_0$ denotes the estimated representation of $x_0$. Here, we employ an attention-based neural network as the core mechanism for the proposed network. This approach has been proven effective in centralized POI recommendation frameworks \cite{2021STAN,10.1145/3477495.3532012}, capturing connections between consecutive check-in activities. Specifically, we use $X_c=[\textbf{e}_{c_1},\textbf{e}_{c_2}
,...,\textbf{e}_{c_M}]\in\mathbb{R}^{M\times d}$ to denote the embedding of the category sequence. Then, we combine the noised target representation with each category embedding $\textbf{e}_{c_m} \in X_c$:
\begin{equation}\label{eq8}
    z_{c_m} = \textbf{e}_{c_m} + \lambda(x_t+\textbf{e}_t),
\end{equation}
where $\textbf{e}_t$ represents the embedding of the corresponding diffusion/reverse step created by following \cite{vaswani2017attention}, and $\lambda$ is a hyperparameter, indicating the level of noise incorporation. Then, the self-attention mechanism is adopted to enhance the revised sequence embedding $Z_c =[z_{c_1},z_{c_2},...,z_{c_M}] \in\mathbb{R}^{M\times d}$. Given three parameters $W_Q$, $W_K$, $W_V \in\mathbb{R}^{d \times d}$, 
the final embedded sequence $E\in\mathbb{R}^{M \times d}$ is defined as follows:
\begin{equation}\label{attn}
  E = Softmax(\frac{QK^T}{\sqrt{d}})\cdot V,
\end{equation}
\noindent{where $Q=Z_cW_Q$, $K=Z_cW_K$, $V=Z_cW_V$}. To this end, the estimated representation is defined as:
\begin{equation}
    \hat{x}_0 = Sum(E^T),
\end{equation}
where $Sum(\cdot)$ is the sum of the last dimension. Then, we utilize the cross-entropy loss for model optimization:
\begin{equation}
  \mathcal{L}_{CE}(\hat{x}_0,x_0) = -\left(log\sigma(\hat{x}^T_0 \cdot x_0)-\frac{1}{|Y^-|}\sum_{{e}_n \in Y^-} log\sigma(\hat{x}^T_0 \cdot \textbf{e}_n)\right),
\end{equation}
where the symbol $\cdot$ denotes the inner product, $Y^-$ consists of multiple negative embeddings for each positive sample where $\textbf{e}_n \neq x_0$, and $\sigma(\cdot)$ is the sigmoid function. 

\subsection{Region-Specific Models}
As of now, we have developed a proficiently trained global network $\Theta_g$. This model is set to undergo further modifications to cater to POI recommendations. A tailored model is established for each region on the corresponding edge server, driven by two key factors. Firstly, region-specific attributes, such as POI-level details and precise geographical data, are potentially redundant or even disruptive when applied outside their respective regions. Secondly, considering that the region-specific model is intended for deployment on user devices, the storage of embeddings for all POIs, rather than just those within a specific region, poses an unnecessary load on the device's resources.

For each region $r$, after receiving the pre-trained global network $\Theta_g$, the edge server freezes its parameters and injects trainable parameters, mainly containing the yet-to-be-trained embeddings of all POIs within the region $r$. We use $\Theta_r$ to indicate the combined network and $\Theta_r^{'}$ to denote the trainable parameters. Given that each POI is linked to a specific category and representative category embeddings are already established, we initialize each POI embedding with its corresponding category embedding. This strategy infuses category-level insights and contributes to a more efficient training process. The novel structure ensures that the frozen parameters maintain the integrity and performance of the global model, which has been trained on a vast dataset. Simultaneously, the trainable parameters leverage this robust foundation to adapt flexibly to the unique characteristics and requirements of different regions.

The network $\Theta_r$, similar to the global model, undergoes training via a diffusion process. Specifically, given a POI sequence $\mathcal{X}=\{p_1, p_2,...,p_{M}\}$ and the target POI $p_{M+1}$, the diffusion algorithm add noise to the target POI embedding $\textbf{e}_{M+1}$, also denoted as $x_0$, referring to the original representation. This process results in the creation of a noised POI representation $x_t$, where $t$ denotes the specific step in the diffusion process. In this way, valuable samples can be obtained to train the region-specific network $\Theta_r$, which takes the noised target POI representation $x_t$ and the POI sequence $\mathcal{X}$. The output is the estimated target POI representation $\hat{x}_0$, aiming to approximate the true target POI embedding $x_0$. 

The approach utilized by the region-specific network $\Theta_r$ exhibits a nuanced divergence from the global network $\Theta_g$, as detailed in \ref{global attention network}. This variation stems from adaptations in both the underlying task requirements and the architectural design of the network. A pivotal element in this modified approach is the integration of both category and POI embeddings for each POI in the check-in sequence, rather than relying exclusively on category embeddings. This integration is operationalized by revising Equation \ref{eq8} as follows:
\begin{equation}\label{hypsource}
z_{m} = e_{p_m} + \gamma e_{c_m} + \lambda(x_t+d_t).
\end{equation}
Similar to $\lambda$, $\gamma$ serves as a hyperparameter, modulating the influence of category embeddings. This alteration ensures a more comprehensive representation by amalgamating the distinct yet complementary information from both category and POI embeddings. 

An additional modification is to introduce the spatiotemporal correlations of the check-in sequence in its final embedding $E$ obtained by Equation \ref{attn}. Capturing spatiotemporal correlations in POI sequences is essential for delivering personalized and contextually relevant POI recommendations, enabling systems to accurately predict user preferences based on the intricate patterns of their movements and timings. Specifically, we encode the spatiotemporal gaps between two check-ins $p_a$ and $p_b$ via 
$\textbf{e}_{ab}^{\Delta}\in\mathbb{R}^d$: 
\begin{equation}
  \textbf{e}_{ab}^{\Delta}=\Delta_{ab}^s \times \textbf{e}_{\Delta_s} + 
  \Delta_{ab}^t \times \textbf{e}_{\Delta_t} 
\end{equation}
\noindent{where $\textbf{e}_{\Delta_s}$ and $\textbf{e}_{\Delta_t}$ are two unit embeddings to represent a 
specific amount of spatial (e.g., one kilometer) or time (e.g., one hour) difference, $\Delta_{ab}^s$ and $\Delta_{ab}^t$are the 
true spatiotemporal differences of $p_a$ and $p_b$ (e.g., 10 kilometers and 5 hours).} On this basis, the embedding of the trajectory spatiotemporal relation matrix is 
$\Delta\in\mathbb{R}^{M\times M}$:
\begin{equation}
  \Delta= \left[ \begin{array}{cccc}
    e_{11}^{\Delta '} & e_{12}^{\Delta '} & ... & e_{1M}^{\Delta '}\\
    e_{21}^{\Delta '} & e_{22}^{\Delta '} & ... & e_{2M}^{\Delta '}\\
    ... & ... & ... & ...\\
    e_{M1}^{\Delta '} & e_{M2}^{\Delta '} & ... & e^{\Delta '}_{MM}
    \end{array} 
    \right ]
\end{equation}
\noindent{where} $e^{\Delta '}_{ab}$ is the element-wise sum of $\textbf{e}_{ab}^{\Delta}$. To this end, we combine the embedded sequence and spatiotemporal differences by modifying Equation \ref{attn}:
\begin{equation}
  E = Softmax(\frac{QK^T+\Delta}{\sqrt{d}})\cdot V.
\end{equation}
Recall that partial parameters of $\Theta_r$ are frozen, its updates are defined as:
\begin{equation}
    \Theta_r^{'} = \Theta_r^{'} - \gamma \frac{\partial \mathcal{L}_{CE}(\Theta_r(x_t,\mathcal{X}),x_0)}{\partial \Theta_r^{'}},
\end{equation}
where $\gamma$ denotes the learning rate. Please note that we do not perform the reverse phase (inference) on the edge server and $\Theta_r$ is directly distributed to all users within this region.

\subsection{Local Finetuning and Inference}
Although the region-specific model can provide POI recommendations for all users within this region, a limitation arises from the model's bias toward active users within the same region, which can compromise overall performance. To mitigate this, there is an initiative to fine-tune the region-specific model locally using personal data. The most straightforward method would involve updating the entire network with local data. Nevertheless, considering the vast number of parameters in the region-specific model, adapting the entire network is not a feasible option. An alternative approach proposes fine-tuning only a subset of parameters. This method, however, has its limitations, as it leads to a performance bottleneck due to the problematic interplay between stable and variable vectors, which disrupts the original feature representation. Hence, inspired by \cite{10.1145/3447548.3467097}, we adopt patch-learning for on-device fine-tuning. This involves integrating an additional patch model into the region-specific network. This patch model undergoes modifications during the local diffusion phase, enabling it to effectively capture user-specific preferences.

Formally, for each user $u$, we introduce a Multi-Layer Perceptron $\Theta_u$, aimed to modify the reconstructed $\hat{x}_0$ presentation returned by $\Theta_r$:
\begin{equation}
    \hat{x}_0 \leftarrow \Theta_u(\Theta_r(x_t,\mathcal{X})).
\end{equation}
Then, we repeat the region-specific diffusion phase locally with personal data to train the user-specific MLP while freezing all parameters of $\Theta_r$:
\begin{equation}
    \Theta_u = \Theta_u - \gamma \frac{\partial \mathcal{L}_{CE}(\hat{x}_0,x_0)}{\partial \Theta_u},
\end{equation}
where $\gamma$ denotes the learning rate. Note that the design of MLP can be adapted to the capacity of the user device. In this work, the MLP with $3$ hidden layers, all having $d$ units, is utilized. To this end, we have described the whole design of DCPR and its optimization is summarized in Algorithm \ref{alg:A}. 

\begin{algorithm}[t]
  \caption{Cloud-Edge-Device Collaborative Training of DCPR.}
  \label{alg:A}
\begin{algorithmic}[1]

    \STATEx /*Global model - on the cloud server*/
    \STATE Initialize $\Theta_g$;
    \FOR{$(c_{M+1},\mathcal{X}_c)\in\mathcal{D}_g$}
        \REPEAT
            \STATE $x_0 \leftarrow \textbf{e}_{c_{M+1}}$;
            \STATE $t \sim U(0,T)$; 
            \STATE $x_t \leftarrow \sqrt{\overline{\alpha}_t}x_0 + \sqrt{1-\overline{\alpha}_t}\epsilon$, $\epsilon \sim \mathcal{N}(0,I)$;
            \STATE $\hat{x}_0 \leftarrow \Theta_g(x_t,\mathcal{X}_c)$;
            \STATE $\Theta_g\leftarrow \Theta_g - \gamma \frac{\partial \mathcal{L}_{CE}(\hat{x}_0,x_0)}{\partial \Theta_g}$;
        \UNTIL{convergence}
    \ENDFOR
    \STATEx /*Region-specific model - on the edge server*/
    \FOR[in parallel]{$r\in\mathcal{R}$} \hspace{0.5cm}$\triangleright$ Each region has an edge server
        \STATE Receive $\Theta_g$ and initialize $\Theta_r^{'}$;
        \STATE Obtain $\Theta_r$ by combining $\Theta_g$ and $\Theta_r^{'}$; 
        \FOR{$(p_{M+1},\mathcal{X})\in\mathcal{D}_r$}
            \REPEAT
                \STATE $x_0 \leftarrow \textbf{e}_{p_{M+1}}$;
                \STATE $t \sim U(0,T)$; 
                \STATE $x_t \leftarrow \sqrt{\overline{\alpha}_t}x_0 + \sqrt{1-\overline{\alpha}_t}\epsilon$, $\epsilon \sim \mathcal{N}(0,I)$;
                \STATE $\hat{x}_0 \leftarrow \Theta_r(x_t,\mathcal{X})$;
                \STATE $\Theta_r^{'} = \Theta_r^{'} - \gamma \frac{\partial \mathcal{L}_{CE}(\hat{x}_0,x_0)}{\partial \Theta_r^{'}}$;
            \UNTIL{convergence}
        \ENDFOR
    \ENDFOR
    \STATEx /*Personal model - on the user side*/
    \FOR[in parallel]{$u\in\mathcal{U}$}
        \STATE Receive $\Theta_r$ and initialize $\Theta_u$;
        \FOR{$(p_{M+1},\mathcal{X})\in\mathcal{X}_u$}    
            \REPEAT
                \STATE $x_0 \leftarrow \textbf{e}_{p_{M+1}}$;
                \STATE $t \sim U(0,T)$; 
                \STATE $x_t \leftarrow \sqrt{\overline{\alpha}_t}x_0 + \sqrt{1-\overline{\alpha}_t}\epsilon$, $\epsilon \sim \mathcal{N}(0,I)$;
                \STATE $\hat{x}_0 \leftarrow \Theta_u(\Theta_r(x_t,\mathcal{X}))$;
                \STATE $\Theta_u \leftarrow \Theta_u - \gamma \frac{\partial \mathcal{L}_{CE}(\hat{x}_0,x_0)}{\partial \Theta_u}$;
            \UNTIL{convergence}
        \ENDFOR
    \ENDFOR
\end{algorithmic}
\end{algorithm}


With the region-specific model $\Theta_r$ and personal patch model $\Theta_u$, the reverse algorithm, as detailed in Section \ref{sdm}, is capable of producing a ranked list of POIs for the next movement of the user $u$. Intuitively, the reverse phase begins with sampling the fully noised target item $x_T$ from a standard Gaussian distribution $\mathcal{N}(0, I)$. The denoising process then proceeds iteratively, where $x_{t-1}$ is obtained from $x_t$ under the guidance of $\hat{x}_0$ which is returned by $\Theta_u(\Theta_r(x_t,\mathcal{X}_c))$. Once the final target POI representation $x_0$ is reached, we map it into the discrete POI index space for final recommendations. To accomplish this, we first compute the score of each of $H$ candidate POIs by:
\begin{equation}
    \alpha(p_h) = x^T_0 \cdot e_{p_h}.
\end{equation}
Then, we rank all scores in descending for final recommendations.

A primary limitation of the standard diffusion model lies in its sluggish generation speed, attributed to the iterative processing required in the reverse phase from $x_T$ to $x_0$ to produce the final representation. This issue is further exacerbated when this processing is executed on-device, due to the restricted computational resources available. To address this challenge, we adopt and adapt the novel sampling technique introduced by \cite{song2020denoising}. This technique significantly reduces the number of sampling steps, thereby markedly improving the efficiency of the generation process. Specifically, instead of performing the reverse process on all steps from $x_T$ to $x_0$, we only perform it on $\{x_{\mathcal{T}_S},x_{\mathcal{T}_{S-1}},...,x_{\mathcal{T}_{1}},x_{\mathcal{T}_{0}}\}$, where $[\mathcal{T}_{S},\mathcal{T}_{S-1},...,\mathcal{T}_{1},\mathcal{T}_{0}]$ is an 
arithmetic and decreasing sub-sequence of $[T,T-1,...,1,0]$. On this basis, the relationship between ${x_{\mathcal{T}_s}}$ and ${x_{\mathcal{T}_{s-1}}}$ is changed, which is defined as:
\begin{equation}
    x_{\mathcal{T}_{s-1}} = \sqrt{\overline{\alpha}_{\mathcal{T}_{s-1}}}(\frac{x_{\mathcal{T}_s}-\sqrt{1-\overline{\alpha}_t}\hat{x}_0}{\sqrt{\overline\alpha}_{\mathcal{T}_s}}) + \sqrt{1-\overline{\alpha}_{\mathcal{T}_{s-1}}} \hat{x}_0.
\end{equation}
To this end, the reverse step $T_R$ can be set to any positive integer which is less than the maximum diffusion step $T$.

\section{Experiments}
In this section, we perform comprehensive experiments with two real-world datasets to evaluate the effectiveness and efficiency of the proposed DCPR. The comparative analysis includes two categories of baselines: centralized POI recommendation systems and on-device POI recommenders. Specifically, our investigation seeks to address the following research questions:

\noindent{\textbf{RQ1}}: How does the DCPR perform compared with state-of-the-art POI recommendation methods?

\noindent{\textbf{RQ2}}: How efficient (i.e., model size and time complexity) is the proposed DCPR compared with other on-device POI recommenders?

\noindent{\textbf{RQ3}}: Is the proposed DCPR efficiently transferable to new regions and new users?

\noindent{\textbf{RQ4}}: What is the impact of DCPR's key hyperparameters?

\begin{table}[htbp]
\vspace{-1.0em}
  \centering
  \caption{Dataset statistics.}
  \label{table:A}
   \vspace{-1.0em}
    \begin{tabular}{lrr}
          \hline
          & \multicolumn{1}{r}{Foursquare} & \multicolumn{1}{r}{Weeplace} \\
    \hline
    \#users & 7,507     & 4,560 \\
    \hline
    \#POIs & 80,962     & 44,194 \\
    \hline
    \#categories & 436     & 625 \\
    \hline
    \#check-ins & 1,214,631     & 923,600 \\
    \hline
    \#check-ins per user & 161.80   & 202.54 \\
    \hline
    \end{tabular}%
    \vspace{-1.0em}
\end{table}%

\begin{table*}[t]
  \centering
  \caption{Recommendation performance comparison with baselines.}
  \label{table:B}
  \vspace{-1.0em}
    \begin{tabular}{|l|cccc|cccc|}
      \hline
          & \multicolumn{4}{c|}{Foursquare}  & \multicolumn{4}{c|}{Weeplace} \\
          \cline{2-9}
          & \multicolumn{1}{c}{HR@5} & \multicolumn{1}{c}{NDCG@5} & \multicolumn{1}{c}{HR@10} & \multicolumn{1}{c|}{NDCG@10} & \multicolumn{1}{c}{HR@5} & \multicolumn{1}{c}{NDCG@5} & \multicolumn{1}{c}{HR@10} & \multicolumn{1}{c|}{NDCG@10} \\
          \hline
          MF    & 0.0847 &0.0607	&0.0965	&0.0661	&0.1042	&0.0599	&0.1316	&0.0889\\
          LSTM  & 0.1939	&0.1195	&0.2782	&0.1668	&0.2156	&0.1322	&0.3251	&0.1549 \\
          STAN  & 0.2987	&0.1776	&0.4327	&0.2598	&0.3141	&0.1819	&0.4663	&0.2876 \\
          DRAN & 0.3114	&0.1802	&0.4345	&0.2655	&0.3165	&0.1843	&0.4775	&0.2974 \\
          Diff-POI & 0.3228	&0.1840	&0.4585	&0.2838	&0.3281	&0.1921	&0.4933	&0.2994 \\
          LLRec & 0.2648	&0.1447	&0.3549	&0.1884	&0.3008	&0.1839	&0.3751	&0.2329\\
          DCCL  & 0.2679	&0.1486	&0.3723	&0.1969	&0.3118	&0.1868	&0.3925	&0.2353\\
          PREFER  & 0.2858	&0.1746	&0.3723	&0.2251	&0.3009	&0.1783	&0.3914	&0.2367 \\
          DCLR  & 0.3136	&0.1887	&0.4406	&0.2740	&0.3124	&0.1857	&0.4357	&0.2797 \\
          MAC  & 0.3030	&0.1826	&0.4520	&0.2780	&0.3187	&0.1974	&0.4852	&0.2829\\
          DCPR & \textbf{0.3272}	&\textbf{0.1922}	&\textbf{0.4623}	&\textbf{0.2913}	&\textbf{0.3337}	&\textbf{0.1978}	&\textbf{0.5082}	&\textbf{0.3063}\\
          \hline
        \end{tabular}%
        \vspace{-1.0em}
  \label{tab:addlabel}%
\end{table*}%

\subsection{Datasets and Evaluation Protocols}

We adopt two real-world datasets to evaluate our proposed DCPR, namely Foursquare \cite{2020Will} and Weeplace \cite{2013Personalized}. Both datasets include users' check-in histories in the cities of New York, Los Angeles, and Chicago. Additionally, in this work, each city is divided into $5$ regions by applying k-means clustering which is discussed in Section \ref{sec:prelim}. Following \cite{Li2018NextPR,2018Content}, users and POIs with less than 10 interactions are removed. Table \ref{table:A} summarizes the statistics of the two datasets. For each dataset, we derive all category sequences from check-in activities and regard those category sequences as the global training data $\mathcal{D}_g$. Then, $50\%$ of the POI sequences within each region $r$ acts as the corresponding region-specific training data $\mathcal{D}_r$, while the rest is employed for the training, testing and validation of on-device models. 

For evaluation, we adopt the leave-one-out protocol which is widely used in previous works \cite{2018Neural,2019Enhancing,zheng2016keyword}. That is, for each of the on-device check-in sequences, the last check-in POI is for testing, the second last POI is for validation, and all others are for training. It is worth noting that, for each category sequence in $\mathcal{D}_g$, we remove the last two check-in activities for rigorous experiments. In addition, the maximum sequence length is set to 200. For each ground truth, instead of ranking all e-commerce products \cite{2020On}, we only pair it with 200 unvisited and nearest POIs within the same region of the sequence as the candidates for ranking. The rationale is, different from e-commerce products \cite{2020On,qu2021imgagn}, in the scenario of POI recommendations that are location-sensitive, users seldom travel between two POIs consecutively that are far away from each other \cite{long2023decentralized,yuan2023federated,li2021discovering}. 

On this basis, we leverage two ranking metrics, namely Hit Ratio at Rank $k$ (HR@$k$) and Normalized Discounted Cumulative Gain at Rank $k$ (NDCG@$k$) \cite{2007CoFiRank} where HR@$k$ only measures the times that the ground truth is present on the top-$k$ list, while NDCG@$k$ cares whether the ground truth can be ranked as highly as possible.

For hyperparameters, we set the maximum diffusion step $T$ to $1024$, the reverse inference step $T_R$ to $16$, $\gamma$ to $0.7$, $\lambda$ to $0.003$, the learning rate to $0.002$, the dimension size to $64$, the dropout to $0.2$, the batch size to $16$, and the maximum training epoch is set to $200$.

\subsection{Baselines}

We compare DCPR with both the centralized and on-device POI recommenders: 

\noindent\textbf{Centralized POI Recommenders:}
\begin{itemize}
  \item \textbf{MF} \cite{2014GeoMF}: It is a classic centralized POI recommendation system based on user-item matrix factorization. 
  
  \item \textbf{LSTM} \cite{1997Long}: This recurrent neural network can capture short-term and long-term dependencies in sequential data.
  
  \item \textbf{STAN} \cite{2021STAN}: It learns explicit spatiotemporal correlations of check-in trajectories via a bi-attention approach.

  \item \textbf{DRAN} \cite{10.1145/3477495.3532012}: It is a GNN-based method that leverages a disentangled representation-enhanced attention network for next POI recommendation

  \item \textbf{Diff-POI} \cite{Qin_2023}: It is a diffusion-based model that samples from the posterior distribution that reflects the user’s geographical preference.  
\end{itemize} 

\textbf{On-Device POI Recommenders:}
\begin{itemize}    
  \item \textbf{LLRec} \cite{2020Next}: It utilizes the teacher-student training strategy to obtain the compressed model that can be deployed locally.

  \item \textbf{DCCL} \cite{10.1145/3447548.3467097}: It compresses and deployes a well-trained model on-device, which is further finetuned locally with personal data.
  
  \item \textbf{PREFER} \cite{2021PREFER}: This federated POI recommendation paradigm allows the server to collect and aggregate locally trained models, as well as redistribute the federated model.
  
  \item \textbf{DCLR} \cite{long2023decentralized}: This decentralized collaborative learning framework allows locally trained models to share knowledge between homogeneous neighbors by model aggregation.

  \item \textbf{MAC} \cite{long2023model}: It is designed to collaboratively train local models with heterogeneous neighbors by comparing their soft decisions on a public reference dataset.
\end{itemize} 

\subsection{Recommendation Effectiveness (RQ1)}
The performance comparison among all the POI recommenders is summarized in Table \ref{table:B}, where we observe the following findings. LSTM outshines MF on both datasets, owing to its adept handling of sequential check-in activities' short-term and long-term dependencies. Furthermore, STAN, which leverages spatiotemporal correlations in check-in activities with the attention mechanism, both consecutive and non-consecutive, surpasses LSTM in accuracy. Advancing further, DRAN melds a Graph Neural Network (GNN) with an attention mechanism, leading to more refined POI embeddings and thus outperforming STAN in terms of accuracy.

Most notably, Diff-POI, employing the robust generality of the diffusion model, sets a new benchmark for state-of-the-art accuracy in this domain. While these centralized models show prowess, our method remains highly competitive. The centralized models, trained across multiple cities, often grapple with the noise in knowledge transfer between cities, which can detract from their performance. In contrast, our approach excels in personalization in regions and personals to learn more expressive models, thereby offering tailored and efficient recommendations.

In the meantime, DCPR outperforms all on-device POI recommenders on both datasets in terms of all metrics. It begins by pointing out the shortcomings of LLRec, which ranks lowest in terms of performance. This is primarily due to its training process that omits all personal data, leading to a disregard for individual user preferences. In contrast, DCCL attempts to enhance its model by incorporating personal data. However, it still does not reach the pinnacle of accuracy, primarily due to its suboptimal model design. The analysis then shifts focus to collaborative learning frameworks such as PREFER, DCLR, and MAC, acknowledging their notable improvements in accuracy. The proposed DCPR stands out for more accurate recommendations. Furthermore, it surpasses other models in recommendation efficiency, demonstrated through its compact on-device model size and reduced computational time complexity. These aspects, along with their implications, are set to be elaborated and explored in greater detail in the subsequent section.

\begin{figure}
	\includegraphics[width=0.8\linewidth]{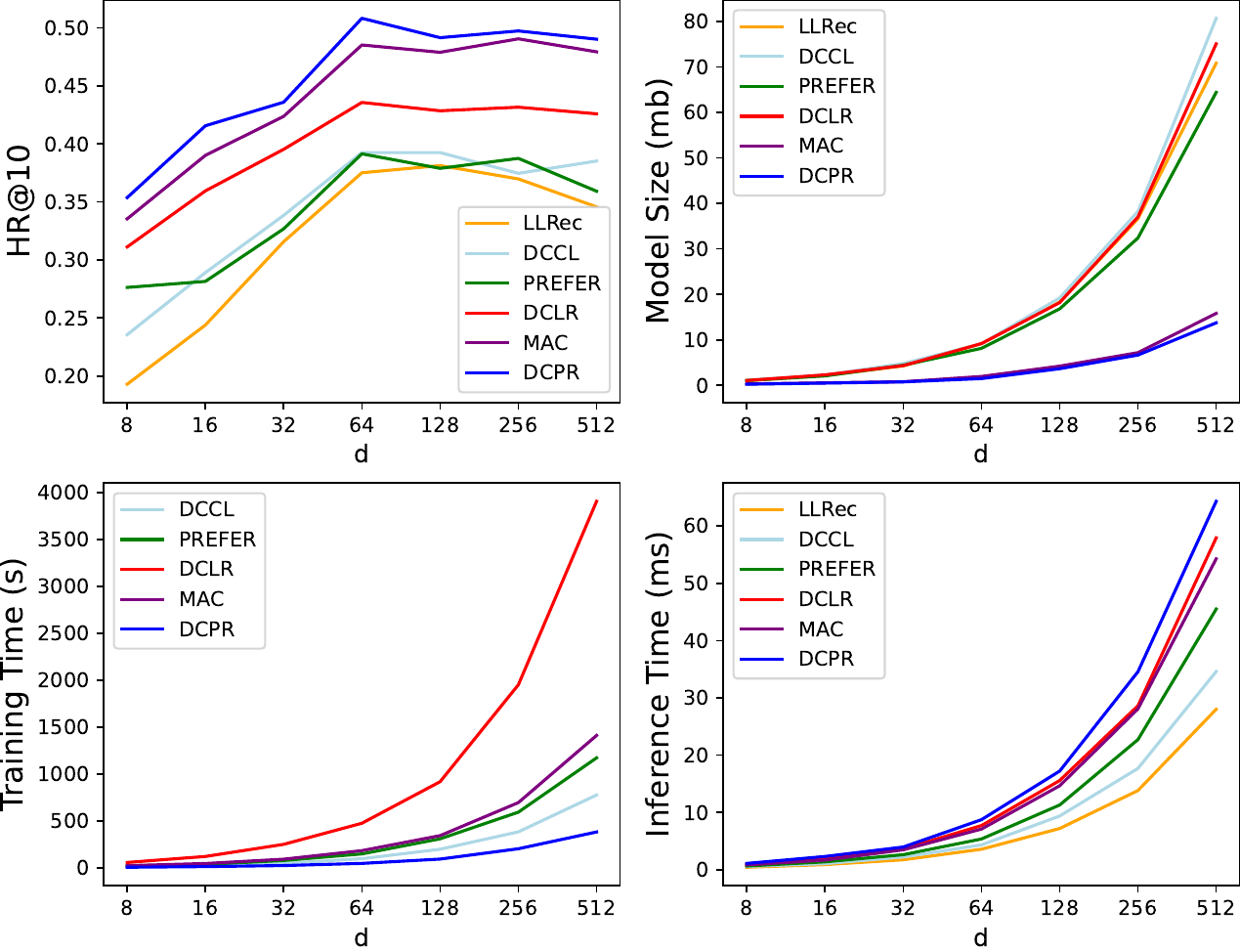}
        \vspace{-1.0em}
	\caption{Recommendation Efficiency.}
	\label{RE} 
\vspace{-1.0em}
\end{figure}

\begin{figure}
	\includegraphics[width=0.8\linewidth]{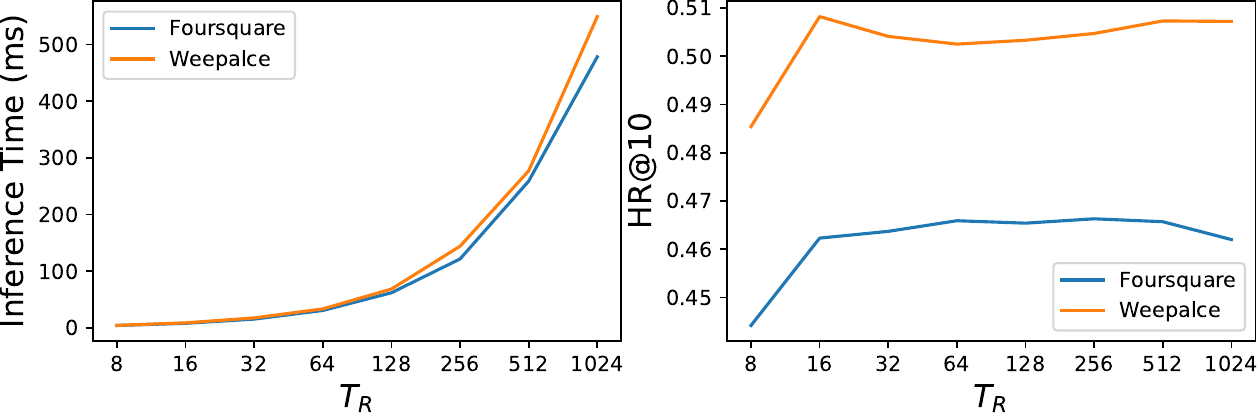}
        \vspace{-1.0em}
	\caption{Effectiveness of Acceleration Strategy.}
	\label{as} 
\vspace{-1.0em}
\end{figure}

\subsection{Recommendation Efficiency (RQ2)}

To assess the recommendation efficiency of the proposed DCPR, concerning all on-device POI recommenders, we record recommendation accuracy (HR@10 on Weeplace), on-device model size (in megabytes), on-device training time (in seconds), and on-device inference time (in milliseconds), for the latent dimensions $d\in\{8,16,32,64,128,256,512\}$. Please note personal models in LLRec are not trained or updated locally, leading to the lack of on-device training time. The summarized results are shown in Figure \ref{RE}.

\subsubsection{On-Device Memory Efficiency}
Here, it is noticeable that the average model sizes of both MAC and DCPR are significantly smaller compared to other on-device recommendation systems. This efficiency is attributed to their design, which allows the user device to store only those POI embeddings that are pertinent.

\subsubsection{On-Device Time Efficiency}\label{tt}
DCPR's efficiency is further highlighted by its minimal reliance on the computational capabilities of local devices, proved by the least on-device training time of DCPR. Regarding inference time, DCPR effectively overcomes the constraints of the standard reverse algorithm, resulting in similar inference time compared with other on-device models. To further prove the effectiveness of the acceleration strategy, we fix the dimension size to $64$, and record the recommendation accuracy and inference time for various $T_R\in\{8, 16, 32, 64, 128, 256, 512, 1024\}$, where $T_R=1024$ means no acceleration is employed. The results are shown in Figure \ref{as}, where we can observe that the acceleration mechanism significantly reduces the inference time while maintaining the recommendation accuracy. More specifically, as $T_R$ increases, there is a generally upward trend in both inference time and accuracy. However, the accuracy converges when $T_R$ exceeds $16$. Meanwhile, the inference time keeps rising with more reverse steps. Thus, $T_R$ is set to $16$, offering high accuracy without excessively prolonging inference time. To conclude, considering the fact that DCPR surpasses all other on-device frameworks in recommendation accuracy, it is a more effective and efficient solution in the landscape of on-device POI recommenders.

\begin{table}[t]
  \centering
  \caption{Recommendation Transferability.}
  \vspace{-1.0em}
  \label{table:trans}
    \begin{tabular}{|l|cc|cc|}
    \hline
          & \multicolumn{2}{c|}{Foursquare} & \multicolumn{2}{c|}{Weeplace} \\
    \cline{2-5}
          & \multicolumn{1}{c}{Time (s)} & \multicolumn{1}{c|}{HR@10} & \multicolumn{1}{c}{Time (s)} & \multicolumn{1}{c|}{HR@10} \\
    \hline
    DCPR & 4143 & 0.4487 & 2824 & 0.4916 \\
    DCPR-T & 5708 & 0.4252 & 4685 & 0.4709 \\
    \hline
    \end{tabular}%
  \label{tab:addlabel}%
\vspace{-0.1cm}
\end{table}%

\subsection{Recommendation Transferability (RQ3)}
As indicated in Section \ref{tt}, the on-device training time for DCPR is significantly shorter compared to other on-device POI recommendation frameworks, while still maintaining high recommendation accuracy. More importantly, the training and updating process for individual user models within DCPR does not interfere with other user models, region-specific models, or the global model, underscoring DCPR's effective adaptability to new users. 

To evaluate DCPR's transferability to new regions, we record the averaged training time (in seconds) of all region-specific models, and recommendation accuracy (HR@10 on Weeplace) after directly applying each of them to personal check-in sequences in the region without local fine-tuning. For comparison, we introduced DCPR-T, a variant where each region-specific model is retrained from scratch, not leveraging the well-trained global model. Similar to DCPR, we record the averaged training time and recommendation accuracy of DCPR-T. The results, shown in Table \ref{table:trans}, reveal that DCPR, with the support of the global model, achieves faster convergence and superior recommendation accuracy compared to DCPR-T. Furthermore, the training of region-specific models does not impact each other or the global model, affirming DCPR's capacity for efficient adaption to new regions. In summary, DCPR's unique cloud-edge-device architecture enables efficient transferability to new users and regions, a feature that distinguishes it from other POI recommendation systems.
\vspace{-0.1cm}
\subsection{Hyperparameter Sensitivity (RQ4)}\label{sec:hyp}
\begin{figure}
	\includegraphics[width=0.8\linewidth]{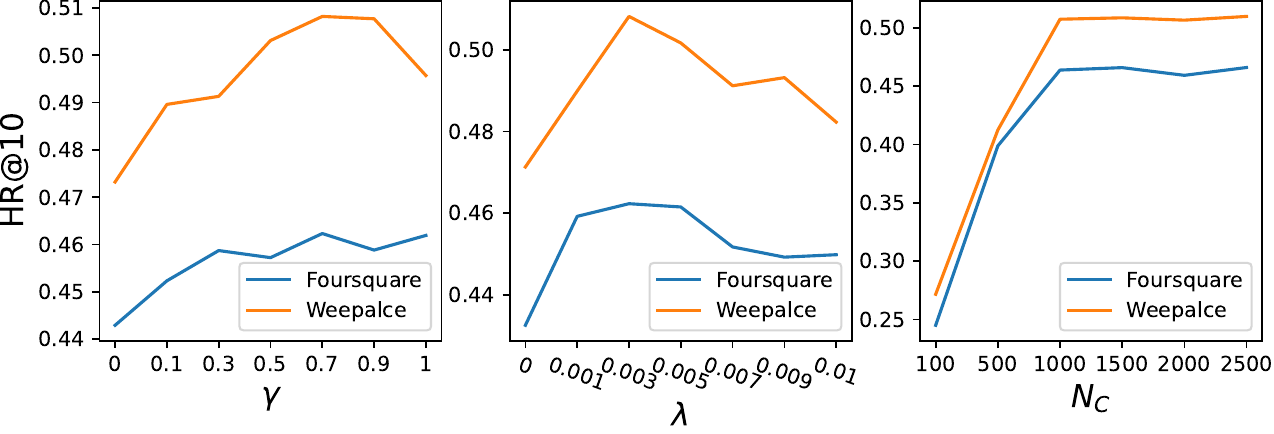}
        \vspace{-1.0em}
	\caption{Hyperparameter Sensitivity.}
	\label{hyp} 
        \vspace{-1.0em}
\end{figure}

In this section, we first illustrate the effect of three hyperparameters on the recommendation accuracy of DCPR including $\gamma$ that controls the injection level of category embedding to POI embedding in Equation \ref{hypsource}, $\lambda$ that controls the noise level added to POI embedding in Equation \ref{hypsource}, and the averaged check-in numbers $N_C$ on the region. The results are shown in Figure \ref{hyp}.

\textbf{Impact of $\gamma$.} We experiment on $\gamma\in\{0,0.1,0.3,0.5,0.7,0.9,1\}$. The lowest accuracy is obtained if all region-specific model is trained solely without well-trained category embeddings in the global model ($\gamma=0$), showing the significance of the cloud-edge-device architecture. Therefore, the recommendation accuracy increases with the increase of $\gamma$. However, the accuracy will decline if knowledge from the frozen category embedding in the global model has an excessive proportion ($\gamma>0.7$).

\textbf{Impact of $\lambda$.} Recommendation accuracy is recorded for $\lambda\in\{0,0.001,0.003,0.005,0.007,0.009,0.01\}$. The best performance is observed when $\gamma=0.003$ for both datasets, highlighting a delicate balance. Excessive noise compromises POI embeddings, while insufficient noise fails to introduce the necessary diversity into POI recommendations. 

\textbf{Impact of $N_C$.} $N_C$is evaluated in $\{100,500,1000,1500,2000,2500\}$. Usually, the recommendation accuracy benefits from higher check-in numbers. In this study, accuracy stabilizes once $N_C$ exceeds $1000$, indicating that the proposed DCPR is capable of delivering high-performance recommendations without requiring an extensive volume of check-ins.

\vspace{-1.0em}

\section{Related Work}
This section reviews recent literature on related areas including centralized models for POI recommendation, on-device frameworks for POI recommendation, and diffusion models.
\vspace{-1.0em}
\subsection{Next POI Recommendation}
To help people discover attractive places by analyzing user-POI interactions, early works mainly focused on matrix factorization \cite{2014GeoMF} and Markov chains \cite{2013Where,zheng2016keyword}. The introduction of recurrent neural network (RNN) models marked a significant advancement, showcasing their ability to understand the spatiotemporal dynamics in POI sequences \cite{ chen2020sequence, 2018DeepMove, chen2020try, yin2015joint, li2021discovering}. Additionally, models employing attentive neural networks \cite{2021STAN, 10.1145/3477495.3531983,chen2019air,yin2015joint} have adopted self-attention mechanisms to meticulously analyze the spatiotemporal context of sequential check-in behaviors. Then, graph neural networks (GNN) based models \cite{li2021discovering,10.1145/3477495.3532012,gao2023graph,gao2023semantic} took a step further by integrating graph-augmented POI sequences, which capitalized on collaborative signals from semantically similar POIs and unveiled sequential trends, thereby outperforming RNN-based approaches in terms of accuracy. Then, Diff-POI \cite{Qin_2023}, by leveraging the powerful generality of the diffusion model, establishes a new standard for cutting-edge accuracy in the field. These approaches, however, predominantly rely on cloud-based infrastructure, which brings the need for substantial cloud computing capabilities. In contrast, DCPR introduces a fast-adapting on-device POI recommendation framework, emphasizing, recommendation accuracy, efficiency, and model transferability.
\subsection{On-Device POI Recommendation}
On-device frameworks effectively address many limitations of cloud-based learning in POI recommendations. Federated learning \cite{2021PREFER}, a key approach in this context, aggregates locally trained models and shares a unified model with users. However, this can result in the long-tail problem, where less active users get subpar recommendations. Some federated POI recommenders \cite{2020xw,2021rao} tackle this by grouping users with similar interests, and decentralized systems \cite{long2023decentralized,long2023model} allow nearby devices to collaborate, enhancing personalization. Despite this, these methods demand extensive device engagement and intra-device communication, raising concerns about privacy and Transferability. An alternative approach \cite{2020Next} is using pre-trained, compressed models on devices with anonymized data for privacy, but this compromises accuracy due to the lack of personalized data and limited model adaptability. Some systems \cite{mairittha2020improving,yan2022device} try to fine-tune with local data, yet they underperform compared to centralized systems. Although this method can quickly adapt to new users, it compromises recommendation quality and struggles to adjust to new regions. Our work introduces a diffusion model-based system, deploying a well-trained model to users for local fine-tuning and achieving high-quality recommendations.

\vspace{-1.0em}
\subsection{Diffusion Models}
Diffusion models have revolutionized generative tasks across fields like computer vision (CV), natural language processing (NLP), and others \cite{avrahami2022blended,li2022diffusion,ramesh2022hierarchical,yang2023diffusion}, with Denoising Diffusion Probabilistic Models (DDPMs) \cite{ho2020denoising} excelling in creating high-quality images. To improve efficiency, Denoising Diffusion Implicit Models (DDIMs) \cite{song2020denoising} reduce sampling steps with minimal impact on diversity. Despite their ability to generate diverse images, controlling the output remains a challenge. Addressing this, text-conditional diffusion models \cite{avrahami2022blended,ramesh2022hierarchical} have emerged, using text encoders to guide image generation by integrating textual and image representations during the diffusion process. These works \cite{li2022diffusion,Qin_2023,li2023diffurec} also extend the diffusion model to sequence-to-sequence tasks, including NLP and sequential recommendations, by training networks to reconstruct targets from noised inputs. Yet, their application in on-device POI recommendation systems is novel. This study pioneers the use of diffusion models for on-device POI recommendations, harnessing their generative capabilities to deliver transferable, accurate, and efficient recommendations, marking a significant advancement in personalized and location-based services.
\vspace{-0.1cm}
\section{Conclusion}
In conclusion, this work has successfully developed and evaluated the Diffusion-Based Cloud-Edge-Device Collaborative Learning (DCPR) framework, a pioneering approach to on-device POI recommendations. By integrating the diffusion model's strengths in generating diverse and distributed representations, DCPR effectively addresses the limitations of existing centralized and collaborative learning systems, particularly in terms of computational efficiency, and the capacity for personalization. Furthermore, the novel cloud-edge-device architecture ensures DCPR's transferability to new regions and users. Experimental results with two real-world datasets have validated DCPR's effectiveness, showcasing its potential to significantly enhance the quality and accessibility of POI recommendations. 
\vspace{-1.0em}
\begin{acks}
This work is supported by Australian Research Council under the streams of Future Fellowship (Grant No. FT210100624), Discovery Early Career Researcher Award (Grants No. DE230101033), Discovery Project (Grants No.DP240101108 and No.DP240101814). 
\end{acks}

\bibliographystyle{ACM-Reference-Format}
\bibliography{myreference}
\end{document}